\documentclass[aps,superscriptaddress,prb,floatfix,twocolumn,amsmath,amssymb,english]{revtex4-1}
\usepackage[colorlinks=true, urlcolor=blue, linkcolor=blue, citecolor=blue, pdftex]{hyperref}
\usepackage{graphicx}
\usepackage{latexsym}
\usepackage{amsmath}
\usepackage{wasysym}
\usepackage{nicefrac}
\usepackage{color}
\usepackage{dcolumn}
\usepackage{bm}
\usepackage{footmisc}
\usepackage{braket}

\newcommand{\dagga}{{\phantom{\dagger}}}

\begin{document}

\title{Variational Monte Carlo study of a two-orbital Hubbard model for the iron pnictides}
\author{Vito Marino}
\affiliation{International School for Advanced Studies (SISSA) and CNR-IOM, Via Bonomea 265, I-34136 Trieste, Italy}
\author{Gabriele Gatti}
\affiliation{Institute for Condensed Matter Physics and Complex Systems, DISAT, Politecnico di Torino, I-10129 Torino, Italy}
\author{Massimo Capone}
\affiliation{International School for Advanced Studies (SISSA) and CNR-IOM, Via Bonomea 265, I-34136 Trieste, Italy}

\author{Luca F. Tocchio}
\affiliation{Institute for Condensed Matter Physics and Complex Systems, DISAT, Politecnico di Torino, I-10129 Torino, Italy}
\date{\today}

\begin{abstract}
We study a two-orbital Hubbard-Kanamori model, which has been originally proposed for iron-based superconductors, using variational Monte Carlo. We span the nonmagnetic sector at both hole-doping and electron-doping, with respect to the half-filled case $n=2$. We report the presence of a superconductive region with a $s^{\pm}$ symmetry only when the half-filled system is in a Mott state, while orbital selectivity is absent. These results are qualitatively different from what was reported in the three-orbital Hubbard-Kanamori model, where a more extended superconductive region was observed with a concomitant development of orbital selectivity, and they are to some extent more reminiscent of the single-band Hubbard model. 
\end{abstract}

\maketitle

\section{Introduction}

The discovery of iron-based superconductors (IBS)~\cite{kamihara2008} opened a new path in the world of high-temperature superconductivity and strongly correlated systems. Since then, a natural question has focused on the comparison with copper-based high-$T_c$ superconductors (cuprates). Similarly to cuprates, which are made of coupled $\textrm{CuO}_2$ layers, IBS are also layered systems made of FeAs layers, where the Fe atoms form a square lattice. Moreover, in both classes of materials, superconductivity emerges by doping a magnetic parent compound, even if the iron materials display a metallic spin-density wave, in contrast with the antiferromagnetic Mott insulator of the cuprates. 

The most fundamental difference is perhaps that, while copper oxides can be described by a single-band model, the low-energy electronic structure of IBS arises from all five $d$ orbitals of iron. This leads to important consequences for the degree and the nature of electronic correlation in these materials~\cite{georges2013,fernandes2022}.  
Many influential research works highlighted the role of the atomic Hund’s exchange coupling in creating a so-called Hund’s metal state~\cite{haule2008,yin2011,georges2013,fanfarillo2015,demedici_explained,isidori,Mravlje,KuglerSrRuO,Ryee,PhysicsToday,CaponeFanfarillo}. 
This state is characterized by orbital-selective correlations, where electrons in different orbitals exhibit varying degrees of correlation\cite{genesis,demedici2014,nica2017,Kugler,kostin2018,capone2018}, and it is associated to enhanced (or even divergent) charge compressibility~\cite{demedici2017,villar2018,chatzi_old,chatzieleftheriou2023}, superconductivity\cite{hu2018,fanfarillo2020}, charge disproportionation\cite{isidori}, and nematicity\cite{yu_nematic,nematic1,fanfarillo2023,valli2026}. 

A dominant fraction of work on Hund's driven correlations is based on Dynamical Mean-Field 
Theory (DMFT)~\cite{georges1996} and related methods, that treat exactly local dynamical correlations while freezing spatial fluctuations. Yet, non-local correlations can play an important role, as expected in scenarios in which superconducting and nematicity are understood in terms of low-energy models based on the exchange of bosons of magnetic origin~\cite{chubukov2012,fernandes2014,chubukov2015,guterding2015,chubukov2016,yao2018,kreisel2017,hu2018,benfatto2018}.

A complementary approach to include accurately non-local correlations and superconductivity is variational Monte Carlo (VMC) with suitable many-body variational wavefunctions. 
This approach has been recently used~\cite{marino2025} to solve a three-orbital Hubbard-Kanamori 
model~\cite{kanamori1963,georges2013,daghofer2010,fanfarillo2020} based on the observation that the Fermi surface states of IBS are made almost exclusively from $d_{xz}$, $d_{yz}$, and $d_{xy}$ orbitals. The results confirm the central role of the Hund's coupling and reveal new phenomena. For large electron-electron repulsion, the model exhibits orbital selectivity in the whole density range between the parent compound with $n=4$ and the half-filled case with $n=3$, with a breaking of the orbital degeneracy between $d_{xz}$ and $d_{yz}$ orbitals. Notably, the orbital with the largest occupation has sizable pairing correlations within the orbital selective 
phase. 
This scenario is not obviously connected with the single-band Hubbard model, where analogous VMC calculations display $d_{x^2-y^2}$-wave superconductivity by doping the half-filled Mott insulator~\cite{misawa2014b,sato2016,tocchio2016}, as well as several of the main features of cuprates. 

In this work, we consider a simpler two-orbital Hubbard-Kanamori model~\cite{raghu2008} that provides a further simplification assuming that the hopping via the $d_{xy}$ orbital could be integrated out and included into next-nearest-neighbor hopping terms within $d_{xz}$ and $d_{yz}$ orbitals. The simplicity of this model led to several early studies~\cite{moreo2009,ran2009,yamase2013,wang2015,tzen2016,dumitrescu2016} despite some clear shortcomings, for instance, the fact that it does not respect all the symmetries of the FeAs planes~\cite{cvetkovic2013}.  More recently, the model has also been shown to display metallic altermagnetism\cite{Giuli2025,Knolle}.

The main goal of this work is to compare this simplified model with the three-orbital counterpart and with the popular single-band model, contributing to the understanding of common threads in strongly correlated superconductivity and in the interplay between band-structure properties and strong correlations.

Our results show that the two-orbital model exhibits superconductivity with extended $s^{\pm}$ symmetry by both hole doping and electron doping with respect to half-filling, as long as the interaction is large enough to make the parent compound a Mott insulator.
The superconductive window is smaller than for the three-band model. In particular, it is very small on the hole-doped site, while it appears to be more extended on the electron-doped side. This asymmetry can be attributed to the particle-hole asymmetry of the bare density of states. Importantly, no evidence for nematicity or orbital selectivity is found.

These results show that, while a tendency towards superconductivity with a meaningful symmetry is realized also for two orbitals, the lack of the $d_{xy}$ one has severe consequences in reducing the effect of the Hund's coupling, which results in the absence of some of its typical fingerprints. Furthermore, in the two-orbital model, superconductivity is only found when the parent system is a Mott insulator, similarly to the familiar picture obtained for the single-band model, despite the different symmetry of pairing.

The paper is organized as follows. First, we introduce the two-orbital model for iron-based superconductors and the variational wave function adopted to investigate superconductivity. We then present the numerical results, highlighting the emergence of a finite superconducting order parameter for both hole and electron doping with the expected $s^{\pm}$ symmetry. Finally, we summarize the main conclusions of the study.

\section{Model and Method}

The two-orbital model proposed in Ref.~\onlinecite{raghu2008}, includes only hopping terms between the $d_{xz}$ and the $d_{yz}$ orbitals, according to the following noninteracting Hamiltonian ${\cal H}_0$, defined in momentum space: 
\begin{equation}\label{eq:TB}
\mathcal{H}_{0} = \sum_k \sum_{\alpha,\beta} \sum_{\sigma} c^\dag_{k,\alpha,\sigma} T_{\alpha,\beta}(k) c^\dagga_{k,\beta,\sigma},
\end{equation}
where $\alpha$ and $\beta$ are orbital indices ($1=xz$, $2=yz$), while $c^\dag_{k,\alpha,\sigma}$ ($c^\dagga_{k,\alpha,\sigma}$) is the fermionic 
operator that creates (annihilates) an electron in orbital $\alpha$, with momentum $k$ and spin $\sigma$. The intra- and inter-orbital hoppings are given as: 
\begin{eqnarray}
&&T_{1,1}(k) = -2t_2 \cos{k_x} - 2t_1 \cos{k_y} - 4t_3 \cos{k_x}\cos{k_y}, \nonumber \\
&&T_{2,2}(k) = -2t_1 \cos{k_x} - 2t_2 \cos{k_y} - 4t_3 \cos{k_x}\cos{k_y}, \nonumber \\
&&T_{1,2}(k) = T_{2,1}(k) = -4t_4 \sin{k_x} \sin{k_y}, \nonumber
\end{eqnarray}
The hopping parameters (in units of eV) are: $t_1=-0.33, t_2=0.385, t_3=-0.234, t_4=-0.26$~\cite{sknepnek2009}.

As customary, we include local Hubbard and Hund's couplings~\cite{kanamori1963,georges2013}:
\begin{eqnarray}
&&\mathcal{H}_{\rm int} = U \sum_{R} \sum_{\alpha} n_{R,\alpha,\uparrow} n_{R,\alpha,\downarrow} \nonumber \\
&&+U^\prime \sum_R \sum_{\alpha \neq \beta} n_{R,\alpha,\uparrow} n_{R,\beta,\downarrow} \nonumber \\
&&+(U^\prime-J) \sum_R \sum_{\alpha > \beta} \sum_{\sigma}  n_{R,\alpha,\sigma} n_{R,\beta,\sigma} \nonumber \\
&&-J \sum_R \sum_{\alpha \neq \beta} c^\dag_{R,\alpha,\uparrow} c^\dagga_{R,\alpha,\downarrow} c^\dag_{R,\beta,\downarrow} c^\dagga_{R,\beta,\uparrow} \nonumber \\
&&+J \sum_R \sum_{\alpha \neq \beta} c^\dag_{R,\alpha,\uparrow} c^\dag_{R,\alpha,\downarrow} c^\dagga_{R,\beta,\downarrow} c^\dagga_{R,\beta,\uparrow},
\end{eqnarray}
where $n_{R,\alpha,\sigma}=c^\dag_{R,\alpha,\sigma} c^\dagga_{R,\alpha,\sigma}$ is the density operator on site $R$, orbital $\alpha$ and spin $\sigma$; $U$ and 
$U^\prime$ are the intra-orbital and inter-orbital Hubbard interactions, respectively, and $J$ is the Hund's coupling. We assume the system to be rotationally invariant, and thus $U'=U-2J$.

\begin{figure}
\includegraphics[width=\columnwidth]{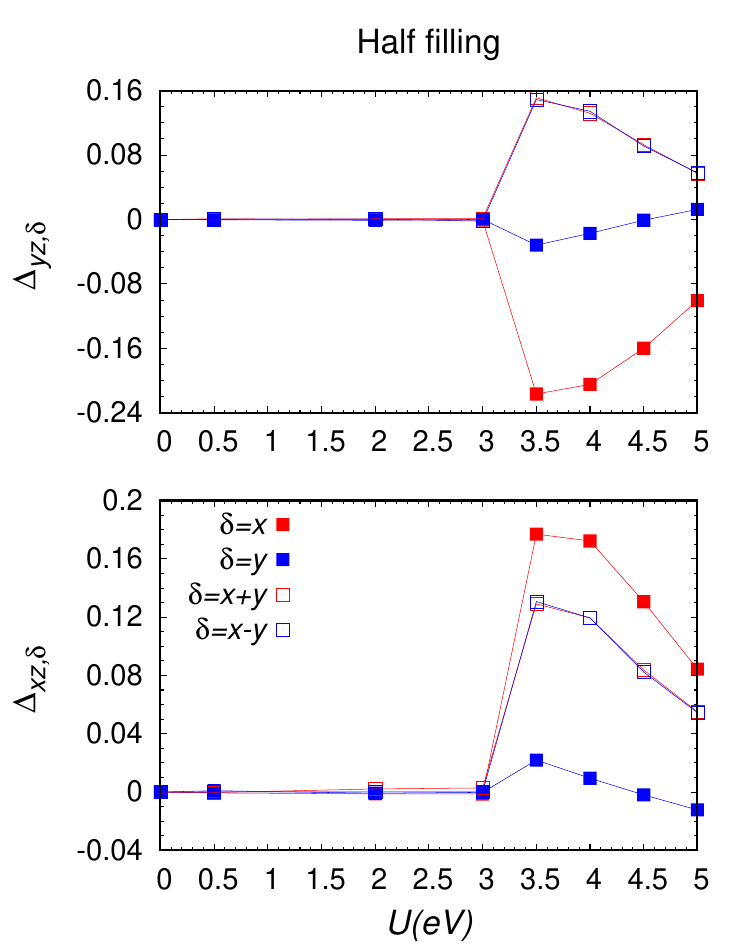}
\caption{\label{fig:BCS-HF}
Upper panel: Optimal intra-orbital BCS parameters in the orbital $d_{yz}$, see  Eq.~(\ref{eq:H_aux}), as a function of $U$ at half filling. Lower panel: Same as in the upper panel, but for the $d_{xz}$ orbital. BCS parameters are computed with the variational approach on a $L=16\times 16$ lattice at $J/U=0.2$.}
\end{figure}

Our results are obtained with the VMC approach, which is based on the definition of correlated variational wave functions, whose parameters and properties can be evaluated within a Monte Carlo 
scheme~\cite{becca2017}. In particular, electron-electron correlation is included by means of multiorbital Jastrow factors~\cite{misawa2014,tocchio2016b,defranco2018,marino2025}, that generalize the Jastrow factors originally proposed for single-orbital Hubbard models~\cite{capello2005}. Jastrow factors are applied on top of an uncorrelated Bardeen-Cooper-Schrieffer (BCS) state, which is appropriate to describe superconductivity:
\begin{equation}\label{eq:psi}
\ket{\Psi} = \mathcal{J}_{c} \mathcal{J}_{s} \ket{\Phi_0}.
\end{equation}
Here, $\mathcal{J}_{c}$ and $\mathcal{J}_{s}$ are density and spin Jastrow factors that are suitable to include correlations within the variational state
\begin{eqnarray}
&\mathcal{J}_{c} = \exp{ \Big( -\frac{1}{2} \sum_{\alpha,\beta} \sum_{R,R^\prime} v_{R,R^\prime}^{\alpha,\beta} n_{R,\alpha} n_{R^\prime,\beta} \Big)} \nonumber \\
&\mathcal{J}_{s} = \exp{ \Big( -\frac{1}{2} \sum_{\alpha \neq \beta} u^{\alpha,\beta} \sum_{R} S_{R,\alpha}^z S_{R,\beta}^z \Big)}
\end{eqnarray}
where $n_{R,\alpha}=n_{R,\alpha,\uparrow}+n_{R,\alpha,\downarrow}$ and $S^z_{i,\alpha}=(n_{R,\alpha,\uparrow}-n_{R,\alpha,\downarrow})/2$ are the total density 
and spin along the $z$ axis on site $R$ and orbital $\alpha$, i.e., $n_{R,\alpha,\sigma}= c^\dag_{R,\alpha,\sigma} c^\dagga_{R,\alpha,\sigma}$ (where 
$c^\dag_{R,\alpha,\sigma}$ and $c^\dagga_{R,\alpha,\sigma}$ are creation and annihilation operators for fermions on site $R$, orbital $\alpha$, and spin 
$\sigma$). The parameters $v_{R,R'}^{\alpha,\beta}$ and $u^{\alpha,\beta}$ are optimized to minimize the variational energy~\cite{sorella2005,becca2017}. 
Notice that the density Jastrow factor includes long-range terms, which were shown to be important to describe the Mott insulator~\cite{capello2005,tocchio2016b,defranco2018}, 
while the spin-spin ones are limited to on-site terms, in order to include the effect of the Hund's coupling. 

The uncorrelated state $\ket{\Phi_0}$ is the ground state 
of an auxiliary Hamiltonian featuring a BCS intra-orbital pairing, in addition to ${\cal H}_0$:
\begin{equation}
 \begin{split}
& \mathcal{H}_{BCS} = \mathcal{H}_{0} - \sum_{R,\alpha,\sigma} \mu_{\alpha} c_{R,\alpha,\sigma}^\dag c_{R,\alpha,\sigma}  \\
    &+ \sum_{R,\alpha,\delta} \left[ \Delta_{\alpha,\delta} \left( c^\dagga_{R,\alpha,\uparrow} c^\dagga_{R+\delta,\alpha,\downarrow}
     - c^\dagga_{R,\alpha,\downarrow} c^\dagga_{R+\delta,\alpha,\uparrow} \right) + \textrm{h.c.}  \right],
\label{eq:H_aux}
\end{split}
\end{equation}
where $\delta=x$, $y$, $x+y$, and $x-y$ indicates nearest and next-nearest neighbors of the site $R$ and $\mu_{\alpha}$ defines the chemical potential 
of orbital $\alpha$. The (singlet) pairing amplitudes $\Delta_{\alpha,\delta}$ and $\mu_{\alpha}$ are also optimized, while the hopping parameters in 
$\mathcal{H}_{0}$ are kept fixed to the bare Hamiltonian. Following previous calculations on three- and five-orbital models, we do not include inter-orbital pairing amplitudes and triplet pairing amplitudes, since they were found to be negligible~\cite{misawa2014,marino2025}. Simulations are done on $L= l \times l$ clusters, with periodic boundary conditions, mainly with $l=12$ and $l=16$.

\begin{figure}
\includegraphics[width=\columnwidth]{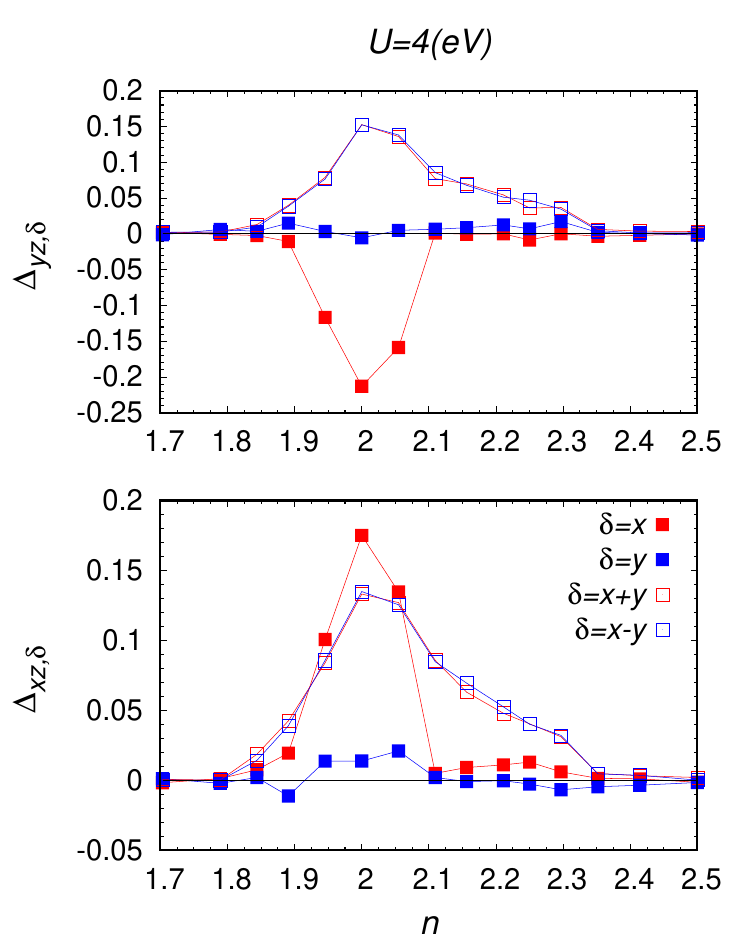}
\caption{\label{fig:BCS-n}
Upper panel: Optimal intra-orbital BCS parameters in the orbital $d_{yz}$, see  Eq.~(\ref{eq:H_aux}), as a function of $n$ at $U=4$ eV. Lower panel: Same as in the upper panel, but for the $d_{xz}$ orbital. BCS parameters are computed with the variational approach on a $L=16\times 16$ lattice at $J/U=0.2$.}
\end{figure}

\section{Results}

In this section, we report on our results for different values of hole and electron doping, with respect to the half-filled case. We have fixed $J/U=0.2$ in all calculations, consistently with Ref.~\onlinecite{marino2025}. 

Within VMC, the presence of superconductivity is
usually associated with the stabilization of BCS parameters in the optimized wave function, see Eq.~(\ref{eq:H_aux}), that defines the uncorrelated state $\ket{\Phi_0}$. Therefore, we start our analysis of superconductivity by showing the optimal (intra-orbital) BCS parameters $\Delta_{\alpha,\delta}$ of Eq.~(\ref{eq:H_aux}) of the main text. 

First, we focus on the half-filled case, which is shown in Fig.~\ref{fig:BCS-HF}. Here, the emergence of finite BCS parameters for values of $U$ that are larger than a critical value of $U_c\sim 3.5$ eV is related to the appearance of a Mott insulator. In this case, the Jastrow factor suppresses the actual pairing correlations, leading to a Mott  insulator~\cite{capello2005,defranco2018}. Similarly to the single-band model, the presence of finite pairing amplitude between the electrons in the Mott insulating state mirrors the resonating valence bond (RVB) picture, originally proposed by Anderson~\cite{anderson1987}, in which superconductivity emerges by doping a Mott insulator that possesses preformed electron pairs.

Then, we consider values of $U$ for which we have a Mott state at half filling, and we present in Fig.~\ref{fig:BCS-n} the evolution of the variational BCS parameters as a function of the electronic density $n$. We observe that, when the electronic density is shifted from half filling, the nearest-neighbor BCS parameters, with $\delta=x$ and $\delta=y$, become rapidly small, while the next-nearest neighbor BCS parameters remain finite and with approximately the same value on the two bonds $\delta=x+y$ and $\delta=x-y$. This behavior is common to both orbitals.

\begin{figure}
\includegraphics[width=\columnwidth]{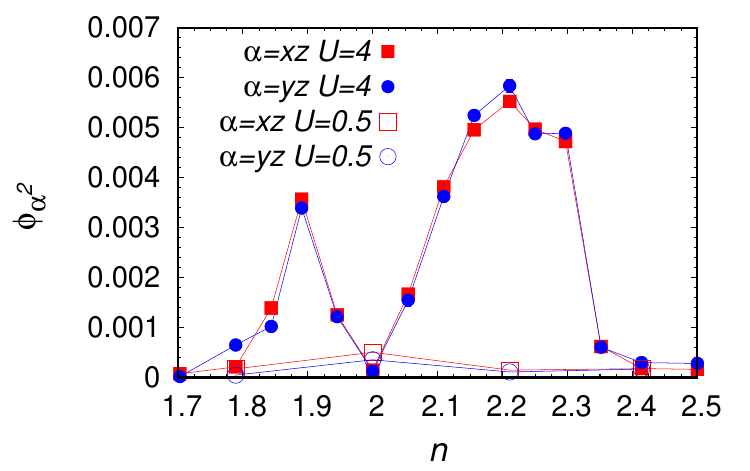}
\caption{\label{fig:pairing}
Square of the superconducting order parameter $\phi_{\alpha}^2$ in each orbital as a function of $n$ at $U=4$ eV (full symbols) and at $U=0.5$ (empty symbols). Data are obtained on the $L=16\times 16$ cluster at $J/U=0.2$.}
\end{figure}

In order to assess the actual presence of a superconducting 
state when doping a Mott insulator, we compute the intra-orbital singlet-singlet pairing correlations $D_{\alpha}(r)$ defined as:

\begin{equation}\begin{split}
\label{eq:pairing}
D_{\alpha}(r)& = \frac{1}{L} \sum_{R} \langle P^\dagga_{R,\alpha} P^\dag_{R+ry,\alpha} \rangle; \\
P^\dagga_{R,\alpha} &= c^\dagga_{R+y+x,\alpha,\downarrow} c^\dagga_{R,\alpha,\uparrow} - c^\dagga_{R+y+x,\alpha,\uparrow} c^\dagga_{R,\alpha,\downarrow}.
\end{split}\end{equation}

The operator $P_{R,\alpha}$ annihilates two electrons on the same orbital at next-nearest-neighbor sites (along the $x+y$ bond). This choice is driven by the fact that the strongest pairing amplitudes in the variational state are those connecting next-nearest neighbor sites, and is further supported in the Appendix. Then, superconductivity is present in orbital $\alpha$ whenever the square of the superconductive order parameter $\phi_{\alpha}^2=\lim_{r \to l/2} D_{\alpha}(r)$ remains finite for large enough lattice sizes $L=l \times l$.

Our results for $\phi_{\alpha}^2$, as defined in Eq.~(\ref{eq:pairing}), are shown in Fig.~\ref{fig:pairing}, as a function of $n$ for $U> U_c$. In particular, we show results for $U=4$ eV, but we also obtain a similar behavior for $U=5$ eV. In both orbitals, we can clearly distinguish a smaller superconductive dome at hole doping and a more extended one at electron doping, separated by a Mott insulating state at half filling. The asymmetry can be connected to a larger bare density of states on the electron-doped side. We compare these results with those obtained for $U=0.5$ eV, representative of the weakly interacting limit. Here, a much smaller order parameter is observed with a maximum at half-filling (which is not Mott insulating). We observe that, within the superconducting domes at $U=4$ eV, our results on the smaller $L=12\times 12$ lattice size (not shown) are almost the same as the ones shown in the figure for $L=16\times 16$, thus indicating that these results are stable with respect to size scaling.

\begin{figure}
\includegraphics[width=\columnwidth]{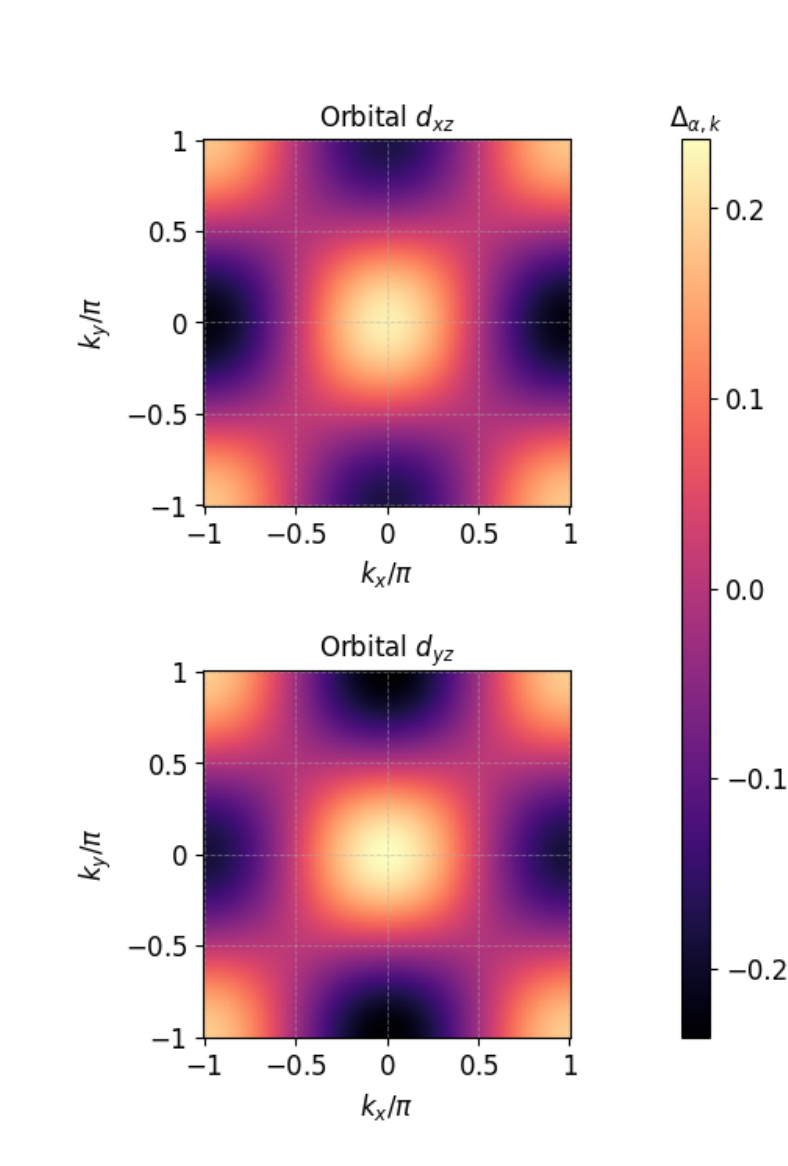}
\caption{\label{fig:BCS-symmetry} Fourier transform of the optimal BCS parameter, as defined in Eq.~(\ref{eq:Delta_Fourier}), for $U=4\,\textrm{eV}$ and $n=2.2$, corresponding to the peak of the superconducting order parameter in the electron-doped regime. Both orbitals, $d_{xz}$ (upper panel) and $d_{yz}$ (lower panel), display an $s^\pm$-wave symmetry.
} 
\end{figure}

The pairing amplitudes are largest on next-nearest neighbor sites in direct space, with the dominant contribution for $\delta=x\pm y$, as shown in Fig.~\ref{fig:BCS-n} and in the Appendix. The symmetry of the BCS parameters is highlighted in $k$-space, where:
\begin{eqnarray}\label{eq:Delta_Fourier}
&& \Delta_{\alpha,k} =2 \left[ \Delta_{\alpha,x} \cos(k_x) + \Delta_{\alpha,y} \cos(k_y) \right. \nonumber \\ 
&&+ \left. \Delta_{\alpha,x-y} \cos(k_x-k_y) + \Delta_{\alpha,x+y} \cos(k_x+k_y) \right].
\end{eqnarray}
Results for both orbitals are reported in Fig.~\ref{fig:BCS-symmetry}, for a value of electron doping that falls within the superconductive dome. Here, $\Delta_{\alpha,k}$ changes sign between the hole pocket around $\Gamma=(0,0)$ and the electron pockets around $X=(\pi,0)$ and $Y=(0,\pi)$, 
resembling the $s^{\pm}$ symmetry.

We remark that superconductivity disappears when we set $J=0$ in our simulations. As already observed in the three-orbital model, the presence of both a finite Coulomb repulsion and a finite Hund's coupling is necessary for the development of superconductivity. This result shall be compared with previous outcomes on a simpler 
tight-binding Hamiltonian $\mathcal{H}_{0}$ with the same diagonal nearest-neighbor hopping for every orbital, where superconductivity was found to be absent at the SU(4) symmetric point with $J=0$~\cite{defranco2018}.   

\begin{figure}
\includegraphics[width=\columnwidth]{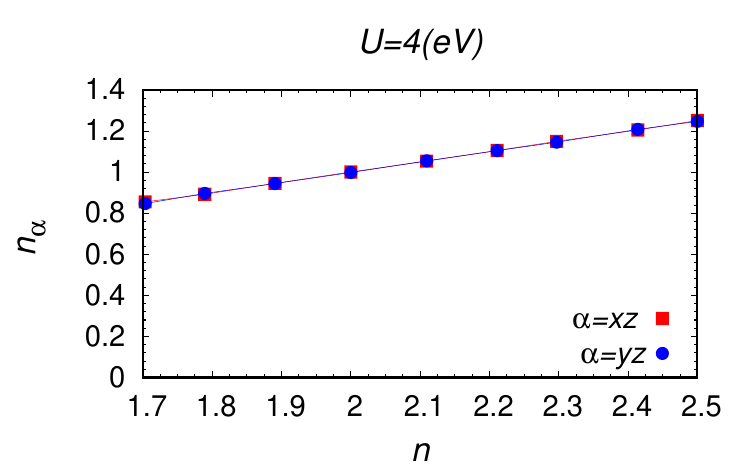}
\caption{\label{fig:density} 
Electronic density per orbital $n_{\alpha}$ as a function of $n$, for $U=4$ eV. Data are obtained for $J/U=0.2$. and they do not show any size dependence when moving from $L=12\times 12$ to $L=16\times 16$ clusters.} 
\end{figure}

We finally comment on the absence of any orbital selectivity in the two-orbital model. In the three-orbital model, orbital selectivity was triggered by a strong differentiation between the occupations of the orbitals $n_{\alpha}=\frac{1}{L}\sum_R n_{R,\alpha}$. Our calculations instead have $n_1 = n_2$ for any parameter we investigated, as shown in  Fig.~\ref{fig:density}, where $n_{\alpha}$ is reported as a function of doping for a value of $U$ where superconductivity is present in the model. In particular, the two orbitals are both at half filling for $n=2$. 

\section{Conclusions} 
We investigated a two-orbital model originally introduced as a possible minimal description of iron-based superconductors using Variational Monte Carlo, incorporating electronic correlations through Jastrow pseudopotentials and superconducting pairing through the auxiliary Hamiltonian.

 For values of $U$ such that the half-filled system ($n=2$) is a Mott insulator, $s^{\pm}$ superconductivity is obtained by doping both with holes and electrons. Electron-doping leads to a wider doping region and larger values of the order parameter consistently with the electron-hole asymmetry of the band structure. On the other hand, the model does not break the symmetry between the two orbitals, which implies that no orbital selectivity or nematic ordering are observed.

Comparing the present results with those of a three-orbital model, we can conclude that the third ($d_{xy}$) orbital included in the model of Ref.~\onlinecite{marino2025} allows for a stronger delocalization of electrons among the orbitals, thereby enhancing the effect of Hund’s coupling and stabilizing the orbital-selective phase with nematic character. 

However, the disappearance of nematicity and orbital selectivity in this simplified model does not destroy also the superconducting phase, which is quite pronounced as long as the half-filled system is in a Mott state. This physics is clearly similar to the single-band Hubbard model, which is in turn considered the basic building block of the cuprate superconductors.
In particular, the present model shows a sizeable pairing amplitude already in the Mott state, in close analogy with the RVB idea put forward by P.W. Anderson. Yet, our results are different from the single-band Hubbard model since they require a finite Hund's coupling to give rise to superconductivity, and they display a different symmetry of the order parameter.

In this regard, our model can be seen as a kind of bridge between two different kinds of strongly correlated superconductors, and it suggests that the very presence of superconductivity with $s_{\pm}$ symmetry in multi-orbital models does not require orbital-selective correlations or nematic ordering. 

\begin{acknowledgments}
We acknowledge useful discussions with Federico Becca. Computational resources were provided by HPC@POLITO (http://www.hpc.polito.it).
\end{acknowledgments}
    
\appendix

\section{Symmetry of the superconductive order parameter}

\begin{figure}[h]
\begin{center}
\includegraphics[width=\columnwidth]{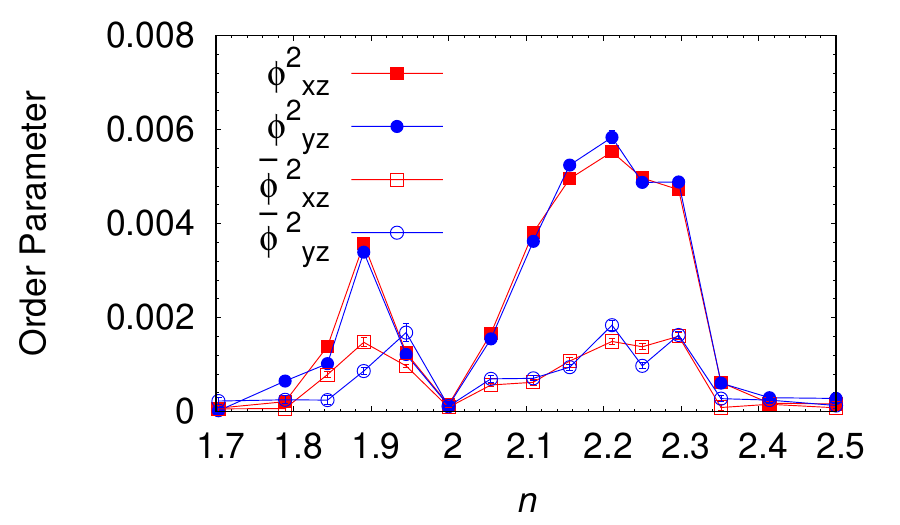}
\caption{\label{fig:sigletorientation} 
Pairing correlations at the maximal distance in the $d_{xz}$ orbital with diagonal singlets, see Eq.~(\ref{eq:pairing1}) ($\phi_{xz}^2$, full red squares),  in the $d_{yz}$ orbitals with diagonal singlets, see Eq.~(\ref{eq:pairing1}) ($\phi_{yz}^2$, full blue circles), in the $d_{xz}$ orbital with horizontal singlets, see Eq.~(\ref{eq:pairing2}) ($\bar{\phi}_{xz}^2$, empty red squares), and in the $d_{yz}$ orbitals with horizontal  singlets, see Eq.~(\ref{eq:pairing2}) ($\bar{\phi}_{yz}^2$, empty blue circles). Results are shown at $U=4$ eV, on the $16\times 16$ cluster for different values of the total electronic density $n$.}
\end{center}
\end{figure}

In order to get further insight into the definition of the superconductive order parameter, we compare, for selected dopings, the values obtained with two different orientations of the singlets. In the first case, introduced in Eq.~(\ref{eq:pairing}) and shown in Fig.~\ref{fig:pairing},  singlets are formed by two electrons at next-nearest neighbor sites, along the $x+y$ bond, i.e.:
\begin{equation}\begin{split}\label{eq:pairing1}
&\phi_{\alpha}^2=\lim_{r \to l/2}  \frac{1}{L} \sum_{R} \langle P^\dagga_{R,\alpha} P^\dag_{R+ry,\alpha} \rangle; \\
& P^\dagga_{R,\alpha} = c^\dagga_{R+y+x,\alpha,\downarrow} c^\dagga_{R,\alpha,\uparrow} - c^\dagga_{R+y+x,\alpha,\uparrow} c^\dagga_{R,\alpha,\downarrow}.
\end{split}\end{equation}
In the second case, singlets are formed by two electrons at nearest neighbor sites, along the $x$ bond, i.e.:
\begin{equation}\begin{split}\label{eq:pairing2}
&\bar{\phi}_{\alpha}^2=\lim_{r \to l/2}  \frac{1}{L} \sum_{R} \langle \bar{P}^\dagga_{R,\alpha} \bar{P}^\dag_{R+ry,\alpha} \rangle; \\
& \bar{P}^{\dagga}_{R,\alpha} = c^\dagga_{R+x,\alpha,\downarrow} c^\dagga_{R,\alpha,\uparrow} - c^\dagga_{R+x,\alpha,\uparrow} c^\dagga_{R,\alpha,\downarrow}.
\end{split}\end{equation}
Results for $U=4$ eV are shown in Fig.~\ref{fig:sigletorientation} and clearly indicate that nearest-neighbor horizontal singlets give a much weaker contribution to superconductivity. It is then justified to determine the presence and the symmetry of superconductivity by focusing on the contribution of singlets formed by electrons at next-nearest neighbor sites along the diagonal bonds.

\bibliography{bibliography}

\begin{thebibliography}{62}%
\makeatletter
\providecommand \@ifxundefined [1]{%
 \@ifx{#1\undefined}
}%
\providecommand \@ifnum [1]{%
 \ifnum #1\expandafter \@firstoftwo
 \else \expandafter \@secondoftwo
 \fi
}%
\providecommand \@ifx [1]{%
 \ifx #1\expandafter \@firstoftwo
 \else \expandafter \@secondoftwo
 \fi
}%
\providecommand \natexlab [1]{#1}%
\providecommand \enquote  [1]{``#1''}%
\providecommand \bibnamefont  [1]{#1}%
\providecommand \bibfnamefont [1]{#1}%
\providecommand \citenamefont [1]{#1}%
\providecommand \href@noop [0]{\@secondoftwo}%
\providecommand \href [0]{\begingroup \@sanitize@url \@href}%
\providecommand \@href[1]{\@@startlink{#1}\@@href}%
\providecommand \@@href[1]{\endgroup#1\@@endlink}%
\providecommand \@sanitize@url [0]{\catcode `\\12\catcode `\$12\catcode
  `\&12\catcode `\#12\catcode `\^12\catcode `\_12\catcode `\%12\relax}%
\providecommand \@@startlink[1]{}%
\providecommand \@@endlink[0]{}%
\providecommand \url  [0]{\begingroup\@sanitize@url \@url }%
\providecommand \@url [1]{\endgroup\@href {#1}{\urlprefix }}%
\providecommand \urlprefix  [0]{URL }%
\providecommand \Eprint [0]{\href }%
\providecommand \doibase [0]{http://dx.doi.org/}%
\providecommand \selectlanguage [0]{\@gobble}%
\providecommand \bibinfo  [0]{\@secondoftwo}%
\providecommand \bibfield  [0]{\@secondoftwo}%
\providecommand \translation [1]{[#1]}%
\providecommand \BibitemOpen [0]{}%
\providecommand \bibitemStop [0]{}%
\providecommand \bibitemNoStop [0]{.\EOS\space}%
\providecommand \EOS [0]{\spacefactor3000\relax}%
\providecommand \BibitemShut  [1]{\csname bibitem#1\endcsname}%
\let\auto@bib@innerbib\@empty
\bibitem [{\citenamefont {Kamihara}\ \emph {et~al.}(2008)\citenamefont
  {Kamihara}, \citenamefont {Watanabe}, \citenamefont {Hirano},\ and\
  \citenamefont {Hosono}}]{kamihara2008}%
  \BibitemOpen
  \bibfield  {author} {\bibinfo {author} {\bibfnamefont {Y.}~\bibnamefont
  {Kamihara}}, \bibinfo {author} {\bibfnamefont {T.}~\bibnamefont {Watanabe}},
  \bibinfo {author} {\bibfnamefont {M.}~\bibnamefont {Hirano}}, \ and\ \bibinfo
  {author} {\bibfnamefont {H.}~\bibnamefont {Hosono}},\ }\href {\doibase
  10.1021/ja800073m} {\bibfield  {journal} {\bibinfo  {journal} {J. Am. Chem.
  Soc.}\ }\textbf {\bibinfo {volume} {130}},\ \bibinfo {pages} {3296} (\bibinfo
  {year} {2008})}\BibitemShut {NoStop}%
\bibitem [{\citenamefont {Georges}\ \emph {et~al.}(2013)\citenamefont
  {Georges}, \citenamefont {Medici},\ and\ \citenamefont
  {Mravlje}}]{georges2013}%
  \BibitemOpen
  \bibfield  {author} {\bibinfo {author} {\bibfnamefont {A.}~\bibnamefont
  {Georges}}, \bibinfo {author} {\bibfnamefont {L.~d.}\ \bibnamefont {Medici}},
  \ and\ \bibinfo {author} {\bibfnamefont {J.}~\bibnamefont {Mravlje}},\ }\href
  {\doibase 10.1146/annurev-conmatphys-020911-125045} {\bibfield  {journal}
  {\bibinfo  {journal} {Annual Review of Condensed Matter Physics}\ }\textbf
  {\bibinfo {volume} {4}},\ \bibinfo {pages} {137} (\bibinfo {year}
  {2013})}\BibitemShut {NoStop}%
\bibitem [{\citenamefont {Fernandes}\ \emph {et~al.}(2022)\citenamefont
  {Fernandes}, \citenamefont {Coldea}, \citenamefont {Ding}, \citenamefont
  {Fisher}, \citenamefont {Hirschfeld},\ and\ \citenamefont
  {Kotliar}}]{fernandes2022}%
  \BibitemOpen
  \bibfield  {author} {\bibinfo {author} {\bibfnamefont {R.~M.}\ \bibnamefont
  {Fernandes}}, \bibinfo {author} {\bibfnamefont {A.~I.}\ \bibnamefont
  {Coldea}}, \bibinfo {author} {\bibfnamefont {H.}~\bibnamefont {Ding}},
  \bibinfo {author} {\bibfnamefont {I.~R.}\ \bibnamefont {Fisher}}, \bibinfo
  {author} {\bibfnamefont {P.~J.}\ \bibnamefont {Hirschfeld}}, \ and\ \bibinfo
  {author} {\bibfnamefont {G.}~\bibnamefont {Kotliar}},\ }\href {\doibase
  10.1038/s41586-021-04073-2} {\bibfield  {journal} {\bibinfo  {journal}
  {Nature}\ }\textbf {\bibinfo {volume} {601}},\ \bibinfo {pages} {35}
  (\bibinfo {year} {2022})}\BibitemShut {NoStop}%
\bibitem [{\citenamefont {Haule}\ \emph {et~al.}(2008)\citenamefont {Haule},
  \citenamefont {Shim},\ and\ \citenamefont {Kotliar}}]{haule2008}%
  \BibitemOpen
  \bibfield  {author} {\bibinfo {author} {\bibfnamefont {K.}~\bibnamefont
  {Haule}}, \bibinfo {author} {\bibfnamefont {J.~H.}\ \bibnamefont {Shim}}, \
  and\ \bibinfo {author} {\bibfnamefont {G.}~\bibnamefont {Kotliar}},\ }\href
  {\doibase 10.1103/PhysRevLett.100.226402} {\bibfield  {journal} {\bibinfo
  {journal} {Phys. Rev. Lett.}\ }\textbf {\bibinfo {volume} {100}},\ \bibinfo
  {pages} {226402} (\bibinfo {year} {2008})}\BibitemShut {NoStop}%
\bibitem [{\citenamefont {Yin}\ \emph {et~al.}(2011)\citenamefont {Yin},
  \citenamefont {Haule},\ and\ \citenamefont {Kotliar}}]{yin2011}%
  \BibitemOpen
  \bibfield  {author} {\bibinfo {author} {\bibfnamefont {Z.~P.}\ \bibnamefont
  {Yin}}, \bibinfo {author} {\bibfnamefont {K.}~\bibnamefont {Haule}}, \ and\
  \bibinfo {author} {\bibfnamefont {G.}~\bibnamefont {Kotliar}},\ }\href
  {\doibase 10.1038/nmat3120} {\bibfield  {journal} {\bibinfo  {journal}
  {Nature Materials}\ }\textbf {\bibinfo {volume} {10}},\ \bibinfo {pages}
  {932} (\bibinfo {year} {2011})}\BibitemShut {NoStop}%
\bibitem [{\citenamefont {Fanfarillo}\ and\ \citenamefont
  {Bascones}(2015)}]{fanfarillo2015}%
  \BibitemOpen
  \bibfield  {author} {\bibinfo {author} {\bibfnamefont {L.}~\bibnamefont
  {Fanfarillo}}\ and\ \bibinfo {author} {\bibfnamefont {E.}~\bibnamefont
  {Bascones}},\ }\href {\doibase 10.1103/PhysRevB.92.075136} {\bibfield
  {journal} {\bibinfo  {journal} {Phys. Rev. B}\ }\textbf {\bibinfo {volume}
  {92}},\ \bibinfo {pages} {075136} (\bibinfo {year} {2015})}\BibitemShut
  {NoStop}%
\bibitem [{\citenamefont {{de' Medici}}(2017)}]{demedici_explained}%
  \BibitemOpen
  \bibfield  {author} {\bibinfo {author} {\bibfnamefont {L.}~\bibnamefont {{de'
  Medici}}},\ }\enquote {\bibinfo {title} {Hund's metals explained},}\ \
  (\bibinfo  {publisher} {E. Pavarini, E. Koch, R. Scalettar, and R. Martin
  (eds.) The Physics of Correlated Insulators, Metals, and Superconductors
  Modeling and Simulation Vol. 7 Forschungszentrum Juelich, ISBN
  978-3-95806-224-5},\ \bibinfo {year} {2017})\BibitemShut {NoStop}%
\bibitem [{\citenamefont {Isidori}\ \emph {et~al.}(2019)\citenamefont
  {Isidori}, \citenamefont {Berovi\ifmmode~\acute{c}\else \'{c}\fi{}},
  \citenamefont {Fanfarillo}, \citenamefont {de' Medici}, \citenamefont
  {Fabrizio},\ and\ \citenamefont {Capone}}]{isidori}%
  \BibitemOpen
  \bibfield  {author} {\bibinfo {author} {\bibfnamefont {A.}~\bibnamefont
  {Isidori}}, \bibinfo {author} {\bibfnamefont {M.}~\bibnamefont
  {Berovi\ifmmode~\acute{c}\else \'{c}\fi{}}}, \bibinfo {author} {\bibfnamefont
  {L.}~\bibnamefont {Fanfarillo}}, \bibinfo {author} {\bibfnamefont
  {L.}~\bibnamefont {de' Medici}}, \bibinfo {author} {\bibfnamefont
  {M.}~\bibnamefont {Fabrizio}}, \ and\ \bibinfo {author} {\bibfnamefont
  {M.}~\bibnamefont {Capone}},\ }\href {\doibase
  10.1103/PhysRevLett.122.186401} {\bibfield  {journal} {\bibinfo  {journal}
  {Phys. Rev. Lett.}\ }\textbf {\bibinfo {volume} {122}},\ \bibinfo {pages}
  {186401} (\bibinfo {year} {2019})}\BibitemShut {NoStop}%
\bibitem [{\citenamefont {Mravlje}\ \emph {et~al.}(2011)\citenamefont
  {Mravlje}, \citenamefont {Aichhorn}, \citenamefont {Miyake}, \citenamefont
  {Haule}, \citenamefont {Kotliar},\ and\ \citenamefont {Georges}}]{Mravlje}%
  \BibitemOpen
  \bibfield  {author} {\bibinfo {author} {\bibfnamefont {J.}~\bibnamefont
  {Mravlje}}, \bibinfo {author} {\bibfnamefont {M.}~\bibnamefont {Aichhorn}},
  \bibinfo {author} {\bibfnamefont {T.}~\bibnamefont {Miyake}}, \bibinfo
  {author} {\bibfnamefont {K.}~\bibnamefont {Haule}}, \bibinfo {author}
  {\bibfnamefont {G.}~\bibnamefont {Kotliar}}, \ and\ \bibinfo {author}
  {\bibfnamefont {A.}~\bibnamefont {Georges}},\ }\href {\doibase
  10.1103/PhysRevLett.106.096401} {\bibfield  {journal} {\bibinfo  {journal}
  {Phys. Rev. Lett.}\ }\textbf {\bibinfo {volume} {106}},\ \bibinfo {pages}
  {096401} (\bibinfo {year} {2011})}\BibitemShut {NoStop}%
\bibitem [{\citenamefont {Kugler}\ \emph {et~al.}(2020)\citenamefont {Kugler},
  \citenamefont {Zingl}, \citenamefont {Strand}, \citenamefont {Lee},
  \citenamefont {von Delft},\ and\ \citenamefont {Georges}}]{KuglerSrRuO}%
  \BibitemOpen
  \bibfield  {author} {\bibinfo {author} {\bibfnamefont {F.~B.}\ \bibnamefont
  {Kugler}}, \bibinfo {author} {\bibfnamefont {M.}~\bibnamefont {Zingl}},
  \bibinfo {author} {\bibfnamefont {H.~U.~R.}\ \bibnamefont {Strand}}, \bibinfo
  {author} {\bibfnamefont {S.-S.~B.}\ \bibnamefont {Lee}}, \bibinfo {author}
  {\bibfnamefont {J.}~\bibnamefont {von Delft}}, \ and\ \bibinfo {author}
  {\bibfnamefont {A.}~\bibnamefont {Georges}},\ }\href {\doibase
  10.1103/PhysRevLett.124.016401} {\bibfield  {journal} {\bibinfo  {journal}
  {Phys. Rev. Lett.}\ }\textbf {\bibinfo {volume} {124}},\ \bibinfo {pages}
  {016401} (\bibinfo {year} {2020})}\BibitemShut {NoStop}%
\bibitem [{\citenamefont {Ryee}\ \emph {et~al.}(2021)\citenamefont {Ryee},
  \citenamefont {Han},\ and\ \citenamefont {Choi}}]{Ryee}%
  \BibitemOpen
  \bibfield  {author} {\bibinfo {author} {\bibfnamefont {S.}~\bibnamefont
  {Ryee}}, \bibinfo {author} {\bibfnamefont {M.~J.}\ \bibnamefont {Han}}, \
  and\ \bibinfo {author} {\bibfnamefont {S.}~\bibnamefont {Choi}},\ }\href
  {\doibase 10.1103/PhysRevLett.126.206401} {\bibfield  {journal} {\bibinfo
  {journal} {Phys. Rev. Lett.}\ }\textbf {\bibinfo {volume} {126}},\ \bibinfo
  {pages} {206401} (\bibinfo {year} {2021})}\BibitemShut {NoStop}%
\bibitem [{\citenamefont {Georges}\ and\ \citenamefont
  {Kotliar}(2024)}]{PhysicsToday}%
  \BibitemOpen
  \bibfield  {author} {\bibinfo {author} {\bibfnamefont {A.}~\bibnamefont
  {Georges}}\ and\ \bibinfo {author} {\bibfnamefont {G.}~\bibnamefont
  {Kotliar}},\ }\href {\doibase 10.1063/pt.wqrz.qpjx} {\bibfield  {journal}
  {\bibinfo  {journal} {Physics Today}\ }\textbf {\bibinfo {volume} {77}},\
  \bibinfo {pages} {46} (\bibinfo {year} {2024})}\BibitemShut {NoStop}%
\bibitem [{\citenamefont {Capone}\ and\ \citenamefont
  {Fanfarillo}(2026)}]{CaponeFanfarillo}%
  \BibitemOpen
  \bibfield  {author} {\bibinfo {author} {\bibfnamefont {M.}~\bibnamefont
  {Capone}}\ and\ \bibinfo {author} {\bibfnamefont {L.}~\bibnamefont
  {Fanfarillo}},\ }\href {\doibase 10.1080/00107514.2026.2647572} {\bibfield
  {journal} {\bibinfo  {journal} {Contemporary Physics}\ }\textbf {\bibinfo
  {volume} {67}},\ \bibinfo {pages} {93} (\bibinfo {year} {2026})},\ \Eprint
  {http://arxiv.org/abs/https://doi.org/10.1080/00107514.2026.2647572}
  {https://doi.org/10.1080/00107514.2026.2647572} \BibitemShut {NoStop}%
\bibitem [{\citenamefont {de'Medici}\ \emph {et~al.}(2009)\citenamefont
  {de'Medici}, \citenamefont {Hassan},\ and\ \citenamefont {Capone}}]{genesis}%
  \BibitemOpen
  \bibfield  {author} {\bibinfo {author} {\bibfnamefont {L.}~\bibnamefont
  {de'Medici}}, \bibinfo {author} {\bibfnamefont {S.~R.}\ \bibnamefont
  {Hassan}}, \ and\ \bibinfo {author} {\bibfnamefont {M.}~\bibnamefont
  {Capone}},\ }\href {\doibase 10.1007/s10948-009-0458-9} {\bibfield  {journal}
  {\bibinfo  {journal} {Journal of Superconductivity and Novel Magnetism}\
  }\textbf {\bibinfo {volume} {22}},\ \bibinfo {pages} {535} (\bibinfo {year}
  {2009})}\BibitemShut {NoStop}%
\bibitem [{\citenamefont {de' Medici}\ \emph {et~al.}(2014)\citenamefont {de'
  Medici}, \citenamefont {Giovannetti},\ and\ \citenamefont
  {Capone}}]{demedici2014}%
  \BibitemOpen
  \bibfield  {author} {\bibinfo {author} {\bibfnamefont {L.}~\bibnamefont {de'
  Medici}}, \bibinfo {author} {\bibfnamefont {G.}~\bibnamefont {Giovannetti}},
  \ and\ \bibinfo {author} {\bibfnamefont {M.}~\bibnamefont {Capone}},\ }\href
  {\doibase 10.1103/PhysRevLett.112.177001} {\bibfield  {journal} {\bibinfo
  {journal} {Phys. Rev. Lett.}\ }\textbf {\bibinfo {volume} {112}},\ \bibinfo
  {pages} {177001} (\bibinfo {year} {2014})}\BibitemShut {NoStop}%
\bibitem [{\citenamefont {Nica}\ \emph {et~al.}(2017)\citenamefont {Nica},
  \citenamefont {Yu},\ and\ \citenamefont {Si}}]{nica2017}%
  \BibitemOpen
  \bibfield  {author} {\bibinfo {author} {\bibfnamefont {E.~M.}\ \bibnamefont
  {Nica}}, \bibinfo {author} {\bibfnamefont {R.}~\bibnamefont {Yu}}, \ and\
  \bibinfo {author} {\bibfnamefont {Q.}~\bibnamefont {Si}},\ }\href {\doibase
  10.1038/s41535-017-0027-6} {\bibfield  {journal} {\bibinfo  {journal} {npj
  Quantum Materials}\ }\textbf {\bibinfo {volume} {2}},\ \bibinfo {pages} {24}
  (\bibinfo {year} {2017})}\BibitemShut {NoStop}%
\bibitem [{\citenamefont {Kugler}\ \emph {et~al.}(2019)\citenamefont {Kugler},
  \citenamefont {Lee}, \citenamefont {Weichselbaum}, \citenamefont {Kotliar},\
  and\ \citenamefont {von Delft}}]{Kugler}%
  \BibitemOpen
  \bibfield  {author} {\bibinfo {author} {\bibfnamefont {F.~B.}\ \bibnamefont
  {Kugler}}, \bibinfo {author} {\bibfnamefont {S.-S.~B.}\ \bibnamefont {Lee}},
  \bibinfo {author} {\bibfnamefont {A.}~\bibnamefont {Weichselbaum}}, \bibinfo
  {author} {\bibfnamefont {G.}~\bibnamefont {Kotliar}}, \ and\ \bibinfo
  {author} {\bibfnamefont {J.}~\bibnamefont {von Delft}},\ }\href {\doibase
  10.1103/PhysRevB.100.115159} {\bibfield  {journal} {\bibinfo  {journal}
  {Phys. Rev. B}\ }\textbf {\bibinfo {volume} {100}},\ \bibinfo {pages}
  {115159} (\bibinfo {year} {2019})}\BibitemShut {NoStop}%
\bibitem [{\citenamefont {Kostin}\ \emph {et~al.}(2018)\citenamefont {Kostin},
  \citenamefont {Sprau}, \citenamefont {Kreisel}, \citenamefont {Chong},
  \citenamefont {B{\"o}hmer}, \citenamefont {Canfield}, \citenamefont
  {Hirschfeld}, \citenamefont {Andersen},\ and\ \citenamefont
  {Davis}}]{kostin2018}%
  \BibitemOpen
  \bibfield  {author} {\bibinfo {author} {\bibfnamefont {A.}~\bibnamefont
  {Kostin}}, \bibinfo {author} {\bibfnamefont {P.~O.}\ \bibnamefont {Sprau}},
  \bibinfo {author} {\bibfnamefont {A.}~\bibnamefont {Kreisel}}, \bibinfo
  {author} {\bibfnamefont {Y.~X.}\ \bibnamefont {Chong}}, \bibinfo {author}
  {\bibfnamefont {A.~E.}\ \bibnamefont {B{\"o}hmer}}, \bibinfo {author}
  {\bibfnamefont {P.~C.}\ \bibnamefont {Canfield}}, \bibinfo {author}
  {\bibfnamefont {P.~J.}\ \bibnamefont {Hirschfeld}}, \bibinfo {author}
  {\bibfnamefont {B.~M.}\ \bibnamefont {Andersen}}, \ and\ \bibinfo {author}
  {\bibfnamefont {J.~C.~S.}\ \bibnamefont {Davis}},\ }\href {\doibase
  10.1038/s41563-018-0151-0} {\bibfield  {journal} {\bibinfo  {journal} {Nature
  Materials}\ }\textbf {\bibinfo {volume} {17}},\ \bibinfo {pages} {869}
  (\bibinfo {year} {2018})}\BibitemShut {NoStop}%
\bibitem [{\citenamefont {Capone}(2018)}]{capone2018}%
  \BibitemOpen
  \bibfield  {author} {\bibinfo {author} {\bibfnamefont {M.}~\bibnamefont
  {Capone}},\ }\href {\doibase 10.1038/s41563-018-0173-7} {\bibfield  {journal}
  {\bibinfo  {journal} {Nature Materials}\ }\textbf {\bibinfo {volume} {17}},\
  \bibinfo {pages} {851} (\bibinfo {year} {2018})}\BibitemShut {NoStop}%
\bibitem [{\citenamefont {de' Medici}(2017)}]{demedici2017}%
  \BibitemOpen
  \bibfield  {author} {\bibinfo {author} {\bibfnamefont {L.}~\bibnamefont {de'
  Medici}},\ }\href {\doibase 10.1103/PhysRevLett.118.167003} {\bibfield
  {journal} {\bibinfo  {journal} {Phys. Rev. Lett.}\ }\textbf {\bibinfo
  {volume} {118}},\ \bibinfo {pages} {167003} (\bibinfo {year}
  {2017})}\BibitemShut {NoStop}%
\bibitem [{\citenamefont {Villar~Arribi}\ and\ \citenamefont {de'
  Medici}(2018)}]{villar2018}%
  \BibitemOpen
  \bibfield  {author} {\bibinfo {author} {\bibfnamefont {P.}~\bibnamefont
  {Villar~Arribi}}\ and\ \bibinfo {author} {\bibfnamefont {L.}~\bibnamefont
  {de' Medici}},\ }\href {\doibase 10.1103/PhysRevLett.121.197001} {\bibfield
  {journal} {\bibinfo  {journal} {Phys. Rev. Lett.}\ }\textbf {\bibinfo
  {volume} {121}},\ \bibinfo {pages} {197001} (\bibinfo {year}
  {2018})}\BibitemShut {NoStop}%
\bibitem [{\citenamefont {Chatzieleftheriou}\ \emph {et~al.}(2020)\citenamefont
  {Chatzieleftheriou}, \citenamefont {Berovi\ifmmode~\acute{c}\else
  \'{c}\fi{}}, \citenamefont {Villar~Arribi}, \citenamefont {Capone},\ and\
  \citenamefont {de' Medici}}]{chatzi_old}%
  \BibitemOpen
  \bibfield  {author} {\bibinfo {author} {\bibfnamefont {M.}~\bibnamefont
  {Chatzieleftheriou}}, \bibinfo {author} {\bibfnamefont {M.}~\bibnamefont
  {Berovi\ifmmode~\acute{c}\else \'{c}\fi{}}}, \bibinfo {author} {\bibfnamefont
  {P.}~\bibnamefont {Villar~Arribi}}, \bibinfo {author} {\bibfnamefont
  {M.}~\bibnamefont {Capone}}, \ and\ \bibinfo {author} {\bibfnamefont
  {L.}~\bibnamefont {de' Medici}},\ }\href {\doibase
  10.1103/PhysRevB.102.205127} {\bibfield  {journal} {\bibinfo  {journal}
  {Phys. Rev. B}\ }\textbf {\bibinfo {volume} {102}},\ \bibinfo {pages}
  {205127} (\bibinfo {year} {2020})}\BibitemShut {NoStop}%
\bibitem [{\citenamefont {Chatzieleftheriou}\ \emph {et~al.}(2023)\citenamefont
  {Chatzieleftheriou}, \citenamefont {Kowalski}, \citenamefont
  {Berovi\ifmmode~\acute{c}\else \'{c}\fi{}}, \citenamefont {Amaricci},
  \citenamefont {Capone}, \citenamefont {De~Leo}, \citenamefont {Sangiovanni},\
  and\ \citenamefont {de' Medici}}]{chatzieleftheriou2023}%
  \BibitemOpen
  \bibfield  {author} {\bibinfo {author} {\bibfnamefont {M.}~\bibnamefont
  {Chatzieleftheriou}}, \bibinfo {author} {\bibfnamefont {A.}~\bibnamefont
  {Kowalski}}, \bibinfo {author} {\bibfnamefont {M.}~\bibnamefont
  {Berovi\ifmmode~\acute{c}\else \'{c}\fi{}}}, \bibinfo {author} {\bibfnamefont
  {A.}~\bibnamefont {Amaricci}}, \bibinfo {author} {\bibfnamefont
  {M.}~\bibnamefont {Capone}}, \bibinfo {author} {\bibfnamefont
  {L.}~\bibnamefont {De~Leo}}, \bibinfo {author} {\bibfnamefont
  {G.}~\bibnamefont {Sangiovanni}}, \ and\ \bibinfo {author} {\bibfnamefont
  {L.}~\bibnamefont {de' Medici}},\ }\href {\doibase
  10.1103/PhysRevLett.130.066401} {\bibfield  {journal} {\bibinfo  {journal}
  {Phys. Rev. Lett.}\ }\textbf {\bibinfo {volume} {130}},\ \bibinfo {pages}
  {066401} (\bibinfo {year} {2023})}\BibitemShut {NoStop}%
\bibitem [{\citenamefont {Hu}\ \emph {et~al.}(2018)\citenamefont {Hu},
  \citenamefont {Yu}, \citenamefont {Nica}, \citenamefont {Zhu},\ and\
  \citenamefont {Si}}]{hu2018}%
  \BibitemOpen
  \bibfield  {author} {\bibinfo {author} {\bibfnamefont {H.}~\bibnamefont
  {Hu}}, \bibinfo {author} {\bibfnamefont {R.}~\bibnamefont {Yu}}, \bibinfo
  {author} {\bibfnamefont {E.~M.}\ \bibnamefont {Nica}}, \bibinfo {author}
  {\bibfnamefont {J.-X.}\ \bibnamefont {Zhu}}, \ and\ \bibinfo {author}
  {\bibfnamefont {Q.}~\bibnamefont {Si}},\ }\href {\doibase
  10.1103/PhysRevB.98.220503} {\bibfield  {journal} {\bibinfo  {journal} {Phys.
  Rev. B}\ }\textbf {\bibinfo {volume} {98}},\ \bibinfo {pages} {220503}
  (\bibinfo {year} {2018})}\BibitemShut {NoStop}%
\bibitem [{\citenamefont {Fanfarillo}\ \emph {et~al.}(2020)\citenamefont
  {Fanfarillo}, \citenamefont {Valli},\ and\ \citenamefont
  {Capone}}]{fanfarillo2020}%
  \BibitemOpen
  \bibfield  {author} {\bibinfo {author} {\bibfnamefont {L.}~\bibnamefont
  {Fanfarillo}}, \bibinfo {author} {\bibfnamefont {A.}~\bibnamefont {Valli}}, \
  and\ \bibinfo {author} {\bibfnamefont {M.}~\bibnamefont {Capone}},\ }\href
  {\doibase 10.1103/PhysRevLett.125.177001} {\bibfield  {journal} {\bibinfo
  {journal} {Phys. Rev. Lett.}\ }\textbf {\bibinfo {volume} {125}},\ \bibinfo
  {pages} {177001} (\bibinfo {year} {2020})}\BibitemShut {NoStop}%
\bibitem [{\citenamefont {Yu}\ \emph {et~al.}(2018)\citenamefont {Yu},
  \citenamefont {Zhu},\ and\ \citenamefont {Si}}]{yu_nematic}%
  \BibitemOpen
  \bibfield  {author} {\bibinfo {author} {\bibfnamefont {R.}~\bibnamefont
  {Yu}}, \bibinfo {author} {\bibfnamefont {J.-X.}\ \bibnamefont {Zhu}}, \ and\
  \bibinfo {author} {\bibfnamefont {Q.}~\bibnamefont {Si}},\ }\href {\doibase
  10.1103/PhysRevLett.121.227003} {\bibfield  {journal} {\bibinfo  {journal}
  {Phys. Rev. Lett.}\ }\textbf {\bibinfo {volume} {121}},\ \bibinfo {pages}
  {227003} (\bibinfo {year} {2018})}\BibitemShut {NoStop}%
\bibitem [{\citenamefont {Fanfarillo}\ \emph {et~al.}(2017)\citenamefont
  {Fanfarillo}, \citenamefont {Giovannetti}, \citenamefont {Capone},\ and\
  \citenamefont {Bascones}}]{nematic1}%
  \BibitemOpen
  \bibfield  {author} {\bibinfo {author} {\bibfnamefont {L.}~\bibnamefont
  {Fanfarillo}}, \bibinfo {author} {\bibfnamefont {G.}~\bibnamefont
  {Giovannetti}}, \bibinfo {author} {\bibfnamefont {M.}~\bibnamefont {Capone}},
  \ and\ \bibinfo {author} {\bibfnamefont {E.}~\bibnamefont {Bascones}},\
  }\href {\doibase 10.1103/PhysRevB.95.144511} {\bibfield  {journal} {\bibinfo
  {journal} {Phys. Rev. B}\ }\textbf {\bibinfo {volume} {95}},\ \bibinfo
  {pages} {144511} (\bibinfo {year} {2017})}\BibitemShut {NoStop}%
\bibitem [{\citenamefont {Fanfarillo}\ \emph {et~al.}(2023)\citenamefont
  {Fanfarillo}, \citenamefont {Valli},\ and\ \citenamefont
  {Capone}}]{fanfarillo2023}%
  \BibitemOpen
  \bibfield  {author} {\bibinfo {author} {\bibfnamefont {L.}~\bibnamefont
  {Fanfarillo}}, \bibinfo {author} {\bibfnamefont {A.}~\bibnamefont {Valli}}, \
  and\ \bibinfo {author} {\bibfnamefont {M.}~\bibnamefont {Capone}},\ }\href
  {\doibase 10.1103/PhysRevB.107.L081114} {\bibfield  {journal} {\bibinfo
  {journal} {Phys. Rev. B}\ }\textbf {\bibinfo {volume} {107}},\ \bibinfo
  {pages} {L081114} (\bibinfo {year} {2023})}\BibitemShut {NoStop}%
\bibitem [{\citenamefont {Valli}\ and\ \citenamefont
  {Fanfarillo}(2026)}]{valli2026}%
  \BibitemOpen
  \bibfield  {author} {\bibinfo {author} {\bibfnamefont {A.}~\bibnamefont
  {Valli}}\ and\ \bibinfo {author} {\bibfnamefont {L.}~\bibnamefont
  {Fanfarillo}},\ }\href {https://arxiv.org/abs/2603.23314} {\enquote {\bibinfo
  {title} {Correlation-driven enhancement of pairing in a nematic hund's
  metal},}\ } (\bibinfo {year} {2026}),\ \Eprint
  {http://arxiv.org/abs/2603.23314} {arXiv:2603.23314 [cond-mat.supr-con]}
  \BibitemShut {NoStop}%
\bibitem [{\citenamefont {Georges}\ \emph {et~al.}(1996)\citenamefont
  {Georges}, \citenamefont {Kotliar}, \citenamefont {Krauth},\ and\
  \citenamefont {Rozenberg}}]{georges1996}%
  \BibitemOpen
  \bibfield  {author} {\bibinfo {author} {\bibfnamefont {A.}~\bibnamefont
  {Georges}}, \bibinfo {author} {\bibfnamefont {G.}~\bibnamefont {Kotliar}},
  \bibinfo {author} {\bibfnamefont {W.}~\bibnamefont {Krauth}}, \ and\ \bibinfo
  {author} {\bibfnamefont {M.~J.}\ \bibnamefont {Rozenberg}},\ }\href {\doibase
  10.1103/RevModPhys.68.13} {\bibfield  {journal} {\bibinfo  {journal} {Rev.
  Mod. Phys.}\ }\textbf {\bibinfo {volume} {68}},\ \bibinfo {pages} {13}
  (\bibinfo {year} {1996})}\BibitemShut {NoStop}%
\bibitem [{\citenamefont {Chubukov}(2012)}]{chubukov2012}%
  \BibitemOpen
  \bibfield  {author} {\bibinfo {author} {\bibfnamefont {A.}~\bibnamefont
  {Chubukov}},\ }\href {\doibase 10.1146/annurev-conmatphys-020911-125055}
  {\bibfield  {journal} {\bibinfo  {journal} {Annual Review of Condensed Matter
  Physics}\ }\textbf {\bibinfo {volume} {3}},\ \bibinfo {pages} {57} (\bibinfo
  {year} {2012})}\BibitemShut {NoStop}%
\bibitem [{\citenamefont {Fernandes}\ \emph {et~al.}(2014)\citenamefont
  {Fernandes}, \citenamefont {Chubukov},\ and\ \citenamefont
  {Schmalian}}]{fernandes2014}%
  \BibitemOpen
  \bibfield  {author} {\bibinfo {author} {\bibfnamefont {R.~M.}\ \bibnamefont
  {Fernandes}}, \bibinfo {author} {\bibfnamefont {A.~V.}\ \bibnamefont
  {Chubukov}}, \ and\ \bibinfo {author} {\bibfnamefont {J.}~\bibnamefont
  {Schmalian}},\ }\href {\doibase 10.1038/nphys2877} {\bibfield  {journal}
  {\bibinfo  {journal} {Nature Physics}\ }\textbf {\bibinfo {volume} {10}},\
  \bibinfo {pages} {97} (\bibinfo {year} {2014})}\BibitemShut {NoStop}%
\bibitem [{\citenamefont {Chubukov}(2015)}]{chubukov2015}%
  \BibitemOpen
  \bibfield  {author} {\bibinfo {author} {\bibfnamefont {A.}~\bibnamefont
  {Chubukov}},\ }\enquote {\bibinfo {title} {Itinerant electron scenario},}\
  in\ \href {\doibase 10.1007/978-3-319-11254-1_8} {\emph {\bibinfo {booktitle}
  {Iron-Based Superconductivity}}},\ \bibinfo {editor} {edited by\ \bibinfo
  {editor} {\bibfnamefont {P.~D.}\ \bibnamefont {Johnson}}, \bibinfo {editor}
  {\bibfnamefont {G.}~\bibnamefont {Xu}}, \ and\ \bibinfo {editor}
  {\bibfnamefont {W.-G.}\ \bibnamefont {Yin}}}\ (\bibinfo  {publisher}
  {Springer International Publishing},\ \bibinfo {year} {2015})\ pp.\ \bibinfo
  {pages} {255--329}\BibitemShut {NoStop}%
\bibitem [{\citenamefont {Guterding}\ \emph {et~al.}(2015)\citenamefont
  {Guterding}, \citenamefont {Backes}, \citenamefont {Jeschke},\ and\
  \citenamefont {Valent\'{\i}}}]{guterding2015}%
  \BibitemOpen
  \bibfield  {author} {\bibinfo {author} {\bibfnamefont {D.}~\bibnamefont
  {Guterding}}, \bibinfo {author} {\bibfnamefont {S.}~\bibnamefont {Backes}},
  \bibinfo {author} {\bibfnamefont {H.~O.}\ \bibnamefont {Jeschke}}, \ and\
  \bibinfo {author} {\bibfnamefont {R.}~\bibnamefont {Valent\'{\i}}},\ }\href
  {\doibase 10.1103/PhysRevB.91.140503} {\bibfield  {journal} {\bibinfo
  {journal} {Phys. Rev. B}\ }\textbf {\bibinfo {volume} {91}},\ \bibinfo
  {pages} {140503} (\bibinfo {year} {2015})}\BibitemShut {NoStop}%
\bibitem [{\citenamefont {Chubukov}\ \emph {et~al.}(2016)\citenamefont
  {Chubukov}, \citenamefont {Khodas},\ and\ \citenamefont
  {Fernandes}}]{chubukov2016}%
  \BibitemOpen
  \bibfield  {author} {\bibinfo {author} {\bibfnamefont {A.~V.}\ \bibnamefont
  {Chubukov}}, \bibinfo {author} {\bibfnamefont {M.}~\bibnamefont {Khodas}}, \
  and\ \bibinfo {author} {\bibfnamefont {R.~M.}\ \bibnamefont {Fernandes}},\
  }\href {\doibase 10.1103/PhysRevX.6.041045} {\bibfield  {journal} {\bibinfo
  {journal} {Phys. Rev. X}\ }\textbf {\bibinfo {volume} {6}},\ \bibinfo {pages}
  {041045} (\bibinfo {year} {2016})}\BibitemShut {NoStop}%
\bibitem [{\citenamefont {Yao}\ and\ \citenamefont {Li}(2018)}]{yao2018}%
  \BibitemOpen
  \bibfield  {author} {\bibinfo {author} {\bibfnamefont {D.-W.}\ \bibnamefont
  {Yao}}\ and\ \bibinfo {author} {\bibfnamefont {T.}~\bibnamefont {Li}},\
  }\href {\doibase 10.1088/1361-648X/aaec23} {\bibfield  {journal} {\bibinfo
  {journal} {Journal of Physics: Condensed Matter}\ }\textbf {\bibinfo {volume}
  {30}},\ \bibinfo {pages} {495601} (\bibinfo {year} {2018})}\BibitemShut
  {NoStop}%
\bibitem [{\citenamefont {Kreisel}\ \emph {et~al.}(2017)\citenamefont
  {Kreisel}, \citenamefont {Andersen}, \citenamefont {Sprau}, \citenamefont
  {Kostin}, \citenamefont {Davis},\ and\ \citenamefont
  {Hirschfeld}}]{kreisel2017}%
  \BibitemOpen
  \bibfield  {author} {\bibinfo {author} {\bibfnamefont {A.}~\bibnamefont
  {Kreisel}}, \bibinfo {author} {\bibfnamefont {B.~M.}\ \bibnamefont
  {Andersen}}, \bibinfo {author} {\bibfnamefont {P.~O.}\ \bibnamefont {Sprau}},
  \bibinfo {author} {\bibfnamefont {A.}~\bibnamefont {Kostin}}, \bibinfo
  {author} {\bibfnamefont {J.~C.~S.}\ \bibnamefont {Davis}}, \ and\ \bibinfo
  {author} {\bibfnamefont {P.~J.}\ \bibnamefont {Hirschfeld}},\ }\href
  {\doibase 10.1103/PhysRevB.95.174504} {\bibfield  {journal} {\bibinfo
  {journal} {Phys. Rev. B}\ }\textbf {\bibinfo {volume} {95}},\ \bibinfo
  {pages} {174504} (\bibinfo {year} {2017})}\BibitemShut {NoStop}%
\bibitem [{\citenamefont {Benfatto}\ \emph {et~al.}(2018)\citenamefont
  {Benfatto}, \citenamefont {Valenzuela},\ and\ \citenamefont
  {Fanfarillo}}]{benfatto2018}%
  \BibitemOpen
  \bibfield  {author} {\bibinfo {author} {\bibfnamefont {L.}~\bibnamefont
  {Benfatto}}, \bibinfo {author} {\bibfnamefont {B.}~\bibnamefont
  {Valenzuela}}, \ and\ \bibinfo {author} {\bibfnamefont {L.}~\bibnamefont
  {Fanfarillo}},\ }\href {\doibase 10.1038/s41535-018-0129-9} {\bibfield
  {journal} {\bibinfo  {journal} {npj Quantum Materials}\ }\textbf {\bibinfo
  {volume} {3}},\ \bibinfo {pages} {56} (\bibinfo {year} {2018})}\BibitemShut
  {NoStop}%
\bibitem [{\citenamefont {Marino}\ \emph {et~al.}(2025)\citenamefont {Marino},
  \citenamefont {Scazzola}, \citenamefont {Becca}, \citenamefont {Capone},\
  and\ \citenamefont {Tocchio}}]{marino2025}%
  \BibitemOpen
  \bibfield  {author} {\bibinfo {author} {\bibfnamefont {V.}~\bibnamefont
  {Marino}}, \bibinfo {author} {\bibfnamefont {A.}~\bibnamefont {Scazzola}},
  \bibinfo {author} {\bibfnamefont {F.}~\bibnamefont {Becca}}, \bibinfo
  {author} {\bibfnamefont {M.}~\bibnamefont {Capone}}, \ and\ \bibinfo {author}
  {\bibfnamefont {L.~F.}\ \bibnamefont {Tocchio}},\ }\href {\doibase
  10.1103/PhysRevLett.134.196502} {\bibfield  {journal} {\bibinfo  {journal}
  {Phys. Rev. Lett.}\ }\textbf {\bibinfo {volume} {134}},\ \bibinfo {pages}
  {196502} (\bibinfo {year} {2025})}\BibitemShut {NoStop}%
\bibitem [{\citenamefont {Kanamori}(1963)}]{kanamori1963}%
  \BibitemOpen
  \bibfield  {author} {\bibinfo {author} {\bibfnamefont {J.}~\bibnamefont
  {Kanamori}},\ }\href@noop {} {\bibfield  {journal} {\bibinfo  {journal}
  {Prog. Theor. Phys.}\ }\textbf {\bibinfo {volume} {30}},\ \bibinfo {pages}
  {275} (\bibinfo {year} {1963})}\BibitemShut {NoStop}%
\bibitem [{\citenamefont {Daghofer}\ \emph {et~al.}(2010)\citenamefont
  {Daghofer}, \citenamefont {Nicholson}, \citenamefont {Moreo},\ and\
  \citenamefont {Dagotto}}]{daghofer2010}%
  \BibitemOpen
  \bibfield  {author} {\bibinfo {author} {\bibfnamefont {M.}~\bibnamefont
  {Daghofer}}, \bibinfo {author} {\bibfnamefont {A.}~\bibnamefont {Nicholson}},
  \bibinfo {author} {\bibfnamefont {A.}~\bibnamefont {Moreo}}, \ and\ \bibinfo
  {author} {\bibfnamefont {E.}~\bibnamefont {Dagotto}},\ }\href {\doibase
  10.1103/PhysRevB.81.014511} {\bibfield  {journal} {\bibinfo  {journal} {Phys.
  Rev. B}\ }\textbf {\bibinfo {volume} {81}},\ \bibinfo {pages} {014511}
  (\bibinfo {year} {2010})}\BibitemShut {NoStop}%
\bibitem [{\citenamefont {Misawa}\ and\ \citenamefont
  {Imada}(2014{\natexlab{a}})}]{misawa2014b}%
  \BibitemOpen
  \bibfield  {author} {\bibinfo {author} {\bibfnamefont {T.}~\bibnamefont
  {Misawa}}\ and\ \bibinfo {author} {\bibfnamefont {M.}~\bibnamefont {Imada}},\
  }\href {\doibase 10.1103/PhysRevB.90.115137} {\bibfield  {journal} {\bibinfo
  {journal} {Phys. Rev. B}\ }\textbf {\bibinfo {volume} {90}},\ \bibinfo
  {pages} {115137} (\bibinfo {year} {2014}{\natexlab{a}})}\BibitemShut
  {NoStop}%
\bibitem [{\citenamefont {Sato}\ and\ \citenamefont
  {Yokoyama}(2016)}]{sato2016}%
  \BibitemOpen
  \bibfield  {author} {\bibinfo {author} {\bibfnamefont {R.}~\bibnamefont
  {Sato}}\ and\ \bibinfo {author} {\bibfnamefont {H.}~\bibnamefont
  {Yokoyama}},\ }\href {\doibase 10.7566/JPSJ.85.074701} {\bibfield  {journal}
  {\bibinfo  {journal} {Journal of the Physical Society of Japan}\ }\textbf
  {\bibinfo {volume} {85}},\ \bibinfo {pages} {074701} (\bibinfo {year}
  {2016})},\ \Eprint
  {http://arxiv.org/abs/https://doi.org/10.7566/JPSJ.85.074701}
  {https://doi.org/10.7566/JPSJ.85.074701} \BibitemShut {NoStop}%
\bibitem [{\citenamefont {Tocchio}\ \emph
  {et~al.}(2016{\natexlab{a}})\citenamefont {Tocchio}, \citenamefont {Becca},\
  and\ \citenamefont {Sorella}}]{tocchio2016}%
  \BibitemOpen
  \bibfield  {author} {\bibinfo {author} {\bibfnamefont {L.~F.}\ \bibnamefont
  {Tocchio}}, \bibinfo {author} {\bibfnamefont {F.}~\bibnamefont {Becca}}, \
  and\ \bibinfo {author} {\bibfnamefont {S.}~\bibnamefont {Sorella}},\ }\href
  {\doibase 10.1103/PhysRevB.94.195126} {\bibfield  {journal} {\bibinfo
  {journal} {Phys. Rev. B}\ }\textbf {\bibinfo {volume} {94}},\ \bibinfo
  {pages} {195126} (\bibinfo {year} {2016}{\natexlab{a}})}\BibitemShut
  {NoStop}%
\bibitem [{\citenamefont {Raghu}\ \emph {et~al.}(2008)\citenamefont {Raghu},
  \citenamefont {Qi}, \citenamefont {Liu}, \citenamefont {Scalapino},\ and\
  \citenamefont {Zhang}}]{raghu2008}%
  \BibitemOpen
  \bibfield  {author} {\bibinfo {author} {\bibfnamefont {S.}~\bibnamefont
  {Raghu}}, \bibinfo {author} {\bibfnamefont {X.-L.}\ \bibnamefont {Qi}},
  \bibinfo {author} {\bibfnamefont {C.-X.}\ \bibnamefont {Liu}}, \bibinfo
  {author} {\bibfnamefont {D.~J.}\ \bibnamefont {Scalapino}}, \ and\ \bibinfo
  {author} {\bibfnamefont {S.-C.}\ \bibnamefont {Zhang}},\ }\href {\doibase
  10.1103/PhysRevB.77.220503} {\bibfield  {journal} {\bibinfo  {journal} {Phys.
  Rev. B}\ }\textbf {\bibinfo {volume} {77}},\ \bibinfo {pages} {220503}
  (\bibinfo {year} {2008})}\BibitemShut {NoStop}%
\bibitem [{\citenamefont {Moreo}\ \emph {et~al.}(2009)\citenamefont {Moreo},
  \citenamefont {Daghofer}, \citenamefont {Riera},\ and\ \citenamefont
  {Dagotto}}]{moreo2009}%
  \BibitemOpen
  \bibfield  {author} {\bibinfo {author} {\bibfnamefont {A.}~\bibnamefont
  {Moreo}}, \bibinfo {author} {\bibfnamefont {M.}~\bibnamefont {Daghofer}},
  \bibinfo {author} {\bibfnamefont {J.~A.}\ \bibnamefont {Riera}}, \ and\
  \bibinfo {author} {\bibfnamefont {E.}~\bibnamefont {Dagotto}},\ }\href
  {\doibase 10.1103/PhysRevB.79.134502} {\bibfield  {journal} {\bibinfo
  {journal} {Phys. Rev. B}\ }\textbf {\bibinfo {volume} {79}},\ \bibinfo
  {pages} {134502} (\bibinfo {year} {2009})}\BibitemShut {NoStop}%
\bibitem [{\citenamefont {Ran}\ \emph {et~al.}(2009)\citenamefont {Ran},
  \citenamefont {Wang}, \citenamefont {Zhai}, \citenamefont {Vishwanath},\ and\
  \citenamefont {Lee}}]{ran2009}%
  \BibitemOpen
  \bibfield  {author} {\bibinfo {author} {\bibfnamefont {Y.}~\bibnamefont
  {Ran}}, \bibinfo {author} {\bibfnamefont {F.}~\bibnamefont {Wang}}, \bibinfo
  {author} {\bibfnamefont {H.}~\bibnamefont {Zhai}}, \bibinfo {author}
  {\bibfnamefont {A.}~\bibnamefont {Vishwanath}}, \ and\ \bibinfo {author}
  {\bibfnamefont {D.-H.}\ \bibnamefont {Lee}},\ }\href {\doibase
  10.1103/PhysRevB.79.014505} {\bibfield  {journal} {\bibinfo  {journal} {Phys.
  Rev. B}\ }\textbf {\bibinfo {volume} {79}},\ \bibinfo {pages} {014505}
  (\bibinfo {year} {2009})}\BibitemShut {NoStop}%
\bibitem [{\citenamefont {Yamase}\ and\ \citenamefont
  {Zeyher}(2013)}]{yamase2013}%
  \BibitemOpen
  \bibfield  {author} {\bibinfo {author} {\bibfnamefont {H.}~\bibnamefont
  {Yamase}}\ and\ \bibinfo {author} {\bibfnamefont {R.}~\bibnamefont
  {Zeyher}},\ }\href {\doibase 10.1103/PhysRevB.88.180502} {\bibfield
  {journal} {\bibinfo  {journal} {Phys. Rev. B}\ }\textbf {\bibinfo {volume}
  {88}},\ \bibinfo {pages} {180502} (\bibinfo {year} {2013})}\BibitemShut
  {NoStop}%
\bibitem [{\citenamefont {Wang}\ and\ \citenamefont
  {Nevidomskyy}(2015)}]{wang2015}%
  \BibitemOpen
  \bibfield  {author} {\bibinfo {author} {\bibfnamefont {Z.}~\bibnamefont
  {Wang}}\ and\ \bibinfo {author} {\bibfnamefont {A.~H.}\ \bibnamefont
  {Nevidomskyy}},\ }\href {\doibase 10.1088/0953-8984/27/22/225602} {\bibfield
  {journal} {\bibinfo  {journal} {Journal of Physics: Condensed Matter}\
  }\textbf {\bibinfo {volume} {27}},\ \bibinfo {pages} {225602} (\bibinfo
  {year} {2015})}\BibitemShut {NoStop}%
\bibitem [{\citenamefont {Ong}\ \emph {et~al.}(2016)\citenamefont {Ong},
  \citenamefont {Coleman},\ and\ \citenamefont {Schmalian}}]{tzen2016}%
  \BibitemOpen
  \bibfield  {author} {\bibinfo {author} {\bibfnamefont {T.}~\bibnamefont
  {Ong}}, \bibinfo {author} {\bibfnamefont {P.}~\bibnamefont {Coleman}}, \ and\
  \bibinfo {author} {\bibfnamefont {J.}~\bibnamefont {Schmalian}},\ }\href
  {\doibase 10.1073/pnas.1523064113} {\bibfield  {journal} {\bibinfo  {journal}
  {Proceedings of the National Academy of Sciences}\ }\textbf {\bibinfo
  {volume} {113}},\ \bibinfo {pages} {5486} (\bibinfo {year} {2016})},\ \Eprint
  {http://arxiv.org/abs/https://www.pnas.org/doi/pdf/10.1073/pnas.1523064113}
  {https://www.pnas.org/doi/pdf/10.1073/pnas.1523064113} \BibitemShut {NoStop}%
\bibitem [{\citenamefont {Dumitrescu}\ \emph {et~al.}(2016)\citenamefont
  {Dumitrescu}, \citenamefont {Serbyn}, \citenamefont {Scalettar},\ and\
  \citenamefont {Vishwanath}}]{dumitrescu2016}%
  \BibitemOpen
  \bibfield  {author} {\bibinfo {author} {\bibfnamefont {P.~T.}\ \bibnamefont
  {Dumitrescu}}, \bibinfo {author} {\bibfnamefont {M.}~\bibnamefont {Serbyn}},
  \bibinfo {author} {\bibfnamefont {R.~T.}\ \bibnamefont {Scalettar}}, \ and\
  \bibinfo {author} {\bibfnamefont {A.}~\bibnamefont {Vishwanath}},\ }\href
  {\doibase 10.1103/PhysRevB.94.155127} {\bibfield  {journal} {\bibinfo
  {journal} {Phys. Rev. B}\ }\textbf {\bibinfo {volume} {94}},\ \bibinfo
  {pages} {155127} (\bibinfo {year} {2016})}\BibitemShut {NoStop}%
\bibitem [{\citenamefont {Cvetkovic}\ and\ \citenamefont
  {Vafek}(2013)}]{cvetkovic2013}%
  \BibitemOpen
  \bibfield  {author} {\bibinfo {author} {\bibfnamefont {V.}~\bibnamefont
  {Cvetkovic}}\ and\ \bibinfo {author} {\bibfnamefont {O.}~\bibnamefont
  {Vafek}},\ }\href {\doibase 10.1103/PhysRevB.88.134510} {\bibfield  {journal}
  {\bibinfo  {journal} {Phys. Rev. B}\ }\textbf {\bibinfo {volume} {88}},\
  \bibinfo {pages} {134510} (\bibinfo {year} {2013})}\BibitemShut {NoStop}%
\bibitem [{\citenamefont {Giuli}\ \emph {et~al.}(2025)\citenamefont {Giuli},
  \citenamefont {Mejuto-Zaera},\ and\ \citenamefont {Capone}}]{Giuli2025}%
  \BibitemOpen
  \bibfield  {author} {\bibinfo {author} {\bibfnamefont {S.}~\bibnamefont
  {Giuli}}, \bibinfo {author} {\bibfnamefont {C.}~\bibnamefont {Mejuto-Zaera}},
  \ and\ \bibinfo {author} {\bibfnamefont {M.}~\bibnamefont {Capone}},\ }\href
  {\doibase 10.1103/PhysRevB.111.L020401} {\bibfield  {journal} {\bibinfo
  {journal} {Phys. Rev. B}\ }\textbf {\bibinfo {volume} {111}},\ \bibinfo
  {pages} {L020401} (\bibinfo {year} {2025})}\BibitemShut {NoStop}%
\bibitem [{\citenamefont {Leeb}\ \emph {et~al.}(2024)\citenamefont {Leeb},
  \citenamefont {Mook}, \citenamefont {\ifmmode~\check{S}\else
  \v{S}\fi{}mejkal},\ and\ \citenamefont {Knolle}}]{Knolle}%
  \BibitemOpen
  \bibfield  {author} {\bibinfo {author} {\bibfnamefont {V.}~\bibnamefont
  {Leeb}}, \bibinfo {author} {\bibfnamefont {A.}~\bibnamefont {Mook}}, \bibinfo
  {author} {\bibfnamefont {L.}~\bibnamefont {\ifmmode~\check{S}\else
  \v{S}\fi{}mejkal}}, \ and\ \bibinfo {author} {\bibfnamefont {J.}~\bibnamefont
  {Knolle}},\ }\href {\doibase 10.1103/PhysRevLett.132.236701} {\bibfield
  {journal} {\bibinfo  {journal} {Phys. Rev. Lett.}\ }\textbf {\bibinfo
  {volume} {132}},\ \bibinfo {pages} {236701} (\bibinfo {year}
  {2024})}\BibitemShut {NoStop}%
\bibitem [{\citenamefont {Sknepnek}\ \emph {et~al.}(2009)\citenamefont
  {Sknepnek}, \citenamefont {Samolyuk}, \citenamefont {Lee},\ and\
  \citenamefont {Schmalian}}]{sknepnek2009}%
  \BibitemOpen
  \bibfield  {author} {\bibinfo {author} {\bibfnamefont {R.}~\bibnamefont
  {Sknepnek}}, \bibinfo {author} {\bibfnamefont {G.}~\bibnamefont {Samolyuk}},
  \bibinfo {author} {\bibfnamefont {Y.-b.}\ \bibnamefont {Lee}}, \ and\
  \bibinfo {author} {\bibfnamefont {J.}~\bibnamefont {Schmalian}},\ }\href
  {\doibase 10.1103/PhysRevB.79.054511} {\bibfield  {journal} {\bibinfo
  {journal} {Phys. Rev. B}\ }\textbf {\bibinfo {volume} {79}},\ \bibinfo
  {pages} {054511} (\bibinfo {year} {2009})}\BibitemShut {NoStop}%
\bibitem [{\citenamefont {Becca}\ and\ \citenamefont
  {Sorella}(2017)}]{becca2017}%
  \BibitemOpen
  \bibfield  {author} {\bibinfo {author} {\bibfnamefont {F.}~\bibnamefont
  {Becca}}\ and\ \bibinfo {author} {\bibfnamefont {S.}~\bibnamefont
  {Sorella}},\ }\href {\doibase 10.1017/9781316417041} {\emph {\bibinfo {title}
  {Quantum Monte Carlo Approaches for Correlated Systems`}}}\ (\bibinfo
  {publisher} {Cambridge University Press},\ \bibinfo {year}
  {2017})\BibitemShut {NoStop}%
\bibitem [{\citenamefont {Misawa}\ and\ \citenamefont
  {Imada}(2014{\natexlab{b}})}]{misawa2014}%
  \BibitemOpen
  \bibfield  {author} {\bibinfo {author} {\bibfnamefont {T.}~\bibnamefont
  {Misawa}}\ and\ \bibinfo {author} {\bibfnamefont {M.}~\bibnamefont {Imada}},\
  }\href {\doibase 10.1038/ncomms6738} {\bibfield  {journal} {\bibinfo
  {journal} {Nature Communications}\ }\textbf {\bibinfo {volume} {5}},\
  \bibinfo {pages} {5738} (\bibinfo {year} {2014}{\natexlab{b}})}\BibitemShut
  {NoStop}%
\bibitem [{\citenamefont {Tocchio}\ \emph
  {et~al.}(2016{\natexlab{b}})\citenamefont {Tocchio}, \citenamefont
  {Arrigoni}, \citenamefont {Sorella},\ and\ \citenamefont
  {Becca}}]{tocchio2016b}%
  \BibitemOpen
  \bibfield  {author} {\bibinfo {author} {\bibfnamefont {L.~F.}\ \bibnamefont
  {Tocchio}}, \bibinfo {author} {\bibfnamefont {F.}~\bibnamefont {Arrigoni}},
  \bibinfo {author} {\bibfnamefont {S.}~\bibnamefont {Sorella}}, \ and\
  \bibinfo {author} {\bibfnamefont {F.}~\bibnamefont {Becca}},\ }\href
  {\doibase 10.1088/0953-8984/28/10/105602} {\bibfield  {journal} {\bibinfo
  {journal} {Journal of Physics: Condensed Matter}\ }\textbf {\bibinfo {volume}
  {28}},\ \bibinfo {pages} {105602} (\bibinfo {year}
  {2016}{\natexlab{b}})}\BibitemShut {NoStop}%
\bibitem [{\citenamefont {De~Franco}\ \emph {et~al.}(2018)\citenamefont
  {De~Franco}, \citenamefont {Tocchio},\ and\ \citenamefont
  {Becca}}]{defranco2018}%
  \BibitemOpen
  \bibfield  {author} {\bibinfo {author} {\bibfnamefont {C.}~\bibnamefont
  {De~Franco}}, \bibinfo {author} {\bibfnamefont {L.~F.}\ \bibnamefont
  {Tocchio}}, \ and\ \bibinfo {author} {\bibfnamefont {F.}~\bibnamefont
  {Becca}},\ }\href {\doibase 10.1103/PhysRevB.98.075117} {\bibfield  {journal}
  {\bibinfo  {journal} {Phys. Rev. B}\ }\textbf {\bibinfo {volume} {98}},\
  \bibinfo {pages} {075117} (\bibinfo {year} {2018})}\BibitemShut {NoStop}%
\bibitem [{\citenamefont {Capello}\ \emph {et~al.}(2005)\citenamefont
  {Capello}, \citenamefont {Becca}, \citenamefont {Fabrizio}, \citenamefont
  {Sorella},\ and\ \citenamefont {Tosatti}}]{capello2005}%
  \BibitemOpen
  \bibfield  {author} {\bibinfo {author} {\bibfnamefont {M.}~\bibnamefont
  {Capello}}, \bibinfo {author} {\bibfnamefont {F.}~\bibnamefont {Becca}},
  \bibinfo {author} {\bibfnamefont {M.}~\bibnamefont {Fabrizio}}, \bibinfo
  {author} {\bibfnamefont {S.}~\bibnamefont {Sorella}}, \ and\ \bibinfo
  {author} {\bibfnamefont {E.}~\bibnamefont {Tosatti}},\ }\href {\doibase
  10.1103/PhysRevLett.94.026406} {\bibfield  {journal} {\bibinfo  {journal}
  {Phys. Rev. Lett.}\ }\textbf {\bibinfo {volume} {94}},\ \bibinfo {pages}
  {026406} (\bibinfo {year} {2005})}\BibitemShut {NoStop}%
\bibitem [{\citenamefont {Sorella}(2005)}]{sorella2005}%
  \BibitemOpen
  \bibfield  {author} {\bibinfo {author} {\bibfnamefont {S.}~\bibnamefont
  {Sorella}},\ }\href {\doibase 10.1103/PhysRevB.71.241103} {\bibfield
  {journal} {\bibinfo  {journal} {Phys. Rev. B}\ }\textbf {\bibinfo {volume}
  {71}},\ \bibinfo {pages} {241103} (\bibinfo {year} {2005})}\BibitemShut
  {NoStop}%
\bibitem [{\citenamefont {Anderson}(1987)}]{anderson1987}%
  \BibitemOpen
  \bibfield  {author} {\bibinfo {author} {\bibfnamefont {P.~W.}\ \bibnamefont
  {Anderson}},\ }\href {\doibase 10.1126/science.235.4793.1196} {\bibfield
  {journal} {\bibinfo  {journal} {Science}\ }\textbf {\bibinfo {volume}
  {235}},\ \bibinfo {pages} {1196} (\bibinfo {year} {1987})},\ \Eprint
  {http://arxiv.org/abs/https://www.science.org/doi/pdf/10.1126/science.235.4793.1196}
  {https://www.science.org/doi/pdf/10.1126/science.235.4793.1196} \BibitemShut
  {NoStop}%
\end{thebibliography}%

\end{document}